\documentclass[sn-chicago]{sn-jnl}

\usepackage{graphicx}%
\usepackage{multirow}%
\usepackage{amsmath,amssymb,amsfonts}%
\usepackage{amsthm}%
\usepackage{mathrsfs}%
\usepackage[title]{appendix}%
\usepackage{xcolor}%
\usepackage{textcomp}%
\usepackage{manyfoot}%
\usepackage{booktabs}%
\usepackage{listings}%
\usepackage[T1]{fontenc}
\usepackage[latin9]{inputenc}
\usepackage{geometry}
\usepackage{amssymb}
\usepackage{graphicx}
\usepackage{esint}
\usepackage[linesnumbered,ruled]{algorithm2e}
\usepackage{tikz}
\usepackage{rotating}
\usepackage{subfig}
\usepackage{pdflscape}

\makeatother

\theoremstyle{thmstyleone}%
\theoremstyle{thmstyletwo}%

\theoremstyle{thmstylethree}%

\begin{document}

\title[Article Title]{Improved MCMC with active subspaces}


\author[1,2]{\fnm{Leonardo} \sur{Ripoli}}\email{l.ripoli@pgr.reading.ac.uk}
\author*[3,4]{\fnm{Richard G.} \sur{Everitt}}\email{richard.everitt@warwick.ac.uk}

\affil*[1]{\orgdiv{Department of Mathematics and Statistics}, \orgname{University of Reading}, \orgaddress{\street{Whiteknights}, \city{Reading}, \postcode{RG6 6AX}, \state{Berkshire}, \country{UK}}}

\affil*[2]{\orgdiv{Centre for Doctoral Training in the Mathematics of Planet Earth}, \orgname{Imperial College London}, \orgaddress{\street{South Kensington Campus}, \city{London}, \postcode{SW7 2AZ}, \country{UK}}}

\affil[3]{\orgdiv{Department of Statistics}, \orgname{University of Warwick}, \orgaddress{\city{Coventry}, \postcode{CV4 7AL}, \state{West Midlands}, \country{UK}}}

\affil[4]{\orgdiv{The Zeeman Institute for Systems Biology \& Infectious Disease Epidemiology Research}, \orgname{University of Warwick}, \orgaddress{\city{Coventry}, \postcode{CV4 7AL}, \state{West Midlands}, \country{UK}}}


\abstract{\citet{constantine_accelerating_2016} introduced a Metropolis-Hastings
(MH) approach that target the \emph{active subspace} of a posterior
distribution: a linearly projected subspace that is informed by the
likelihood. \citet{schuster_exact_2017} refined this approach to
introduce a pseudo-marginal Metropolis-Hastings, integrating out \emph{inactive}
variables through estimating a marginal likelihood at every MH iteration.
In this paper we show empirically that the effectiveness of these
approaches is limited in the case where the linearity assumption is
violated, and suggest a particle marginal Metropolis-Hastings algorithm
as an alternative for this situation. Finally, the high computational
cost of these approaches leads us to consider alternative approaches
to using active subspaces in MCMC that avoid the need to estimate
a marginal likelihood: we introduce Metropolis-within-Gibbs and Metropolis
within particle Gibbs methods that provide a more computationally
efficient use of the active subspace.}

\keywords{Markov chain Monte Carlo, active subspaces, sequential Monte Carlo}



\maketitle

\section{Introduction}

\subsection{Motivation}

In this paper we study inference of the Bayesian posterior distribution
$\pi\left(\theta\right)=p\left(\theta\right)l\left(\theta\right)/Z$
on parameter $\theta\in\mathbb{R}^{d}$, where $p$ is a prior distribution
and $l(\theta):=f(y\mid\theta)$ is a likelihood resulting from data
model $f$. We use $\tilde{\pi}\left(\theta\right)$ to denote the
numerator in this expression, the unnormalised posterior. Monte Carlo
methods are the most widely-used approaches for estimating quantities
of interest derived from the posterior distribution. Markov chain
Monte Carlo (MCMC), importance sampling (IS) and sequential Monte
Carlo (SMC) form the basis of most methods, with the most appropriate
approach being application specific. The computational cost of these
methods all depend on the dimension $d$: both in the cost per iteration
of the methods, and in the number of iterations that are required
in order to achieve a given level of accuracy. The most effective
approach for a particular situation may involve combining different
approaches. One of the most prominent examples involving the methods
already mentioned here is particle MCMC \citep{Andrieu2010c}: the
use of SMC to construct effective MCMC proposal distributions for
latent time series and other models.

In complex models it is common that $\theta$ is not fully identifiable
from the data. \citet{brouwer_underlying_2018} provide a review of
inference techniques that can be used in such situations. In particular,
\citet{constantine_accelerating_2016} introduces the idea of identifying
the subspace of the parameter space that is informed by the likelihood,
the \emph{active subspace} (AS), and using MCMC on this space only.
The advantage of this approach, when it is possible, is that the dimension
of the AS is less than $d$, therefore we expect that our MCMC will
not need to be run for as many iterations as we would require for
a $d$-dimensional space.

In this paper we review the method of \citet{constantine_accelerating_2016},
and its successor in \citet{schuster_exact_2017}, which use a Metropolis-Hastings
algorithm on a linearly projected subspace, using an estimated marginal
likelihood which integrates over the remainder of the space. We spend
the remainder of this section describing these approaches, and discussing
their stengths and limitations, with empirical examples. We then make
the following contributions:
\begin{itemize}
\item a particle marginal Metropolis-Hastings (PMMH) algorithm that yields
a more statistically efficient MCMC than the methods of \citet{constantine_accelerating_2016}
and \citet{schuster_exact_2017} (section \ref{sec:Particle-marginal-MH});
\item Metropolis-within-Gibbs and Metropolis within particle Gibbs methods
that permit computationally efficient use of the active subspace and
perform well in non-linear scenarios (section \ref{sec:Metropolis-within-Gibbs-and-part}).
\end{itemize}

\subsection{Active subspaces and MCMC}

\subsubsection{Accelerating MCMC with active subspaces}

In this section we describe the method introduced by \citet{constantine_accelerating_2016},
which is the first to describe an approach to using active subspaces
in MCMC. We denote the \emph{active variables} by $a$ (in $\mathbb{R}^{d_{a}})$
and \emph{inactive variables} by $i$ (in $\mathbb{R}^{d_{i}})$.
Our aim is to find a reparameterisation of $\theta$ of the form
\begin{equation}
\theta=B_{a}a+B_{i}i,
\end{equation}
such that the marginal distribution of $a$ captures most of the variability
of $\pi$, where $B_{a}\in\mathbb{R}^{d\times d_{a}}$ and $B_{i}\in\mathbb{R}^{d\times d_{i}}$
are such that $\left[B_{a},B_{i}\right]$ is an orthonormal basis
of $\mathbb{R}^{d}$. \citet{constantine_accelerating_2016} finds
this reparameterisation using an eigendecomposition of the Monte Carlo
approximation
\begin{equation}
\frac{1}{N}\sum_{m=1}^{N}\nabla\log l\left(\theta^{m}\right)\nabla\log l\left(\theta^{m}\right)^{T}\label{eq:est_as}
\end{equation}
for $\theta^{i}\sim p$ for $n=1:M$, this being an estimate of the
eigendecomposition of the matrix
\begin{equation}
\int_{\theta}\nabla\log l\left(\theta\right)\nabla\log l\left(\theta\right)^{T}p\left(\theta\right)d\theta.\label{eq:as}
\end{equation}
The eigenvectors of this decomposition with the largest eigenvalues
will indicate the directions in which there is most variation in $\nabla\log l\left(\theta\right)$.
The directions in which there is little variation in $\nabla\log l\left(\theta\right)$
- when the likelihood is flat - are those that we may designate as
inactive variables.

\citet{constantine_accelerating_2016} runs a Metropolis-Hastings
(MH) algorithm on the $a$ variables as follows. At the $m$th iteration
of the algorithm, the proposal $\ensuremath{a^{*m}\sim q_{a}\left(\cdot\mid a^{m-1}\right)}$
is made (where $a^{m-1}$ is the current state of the chain), and
this proposal is accepted with probability
\[
\alpha_{a}^{m}=1\wedge\frac{p_{a}\left(a^{*m}\right)\bar{l}_{a}\left(a^{*m}\right)}{p_{a}\left(a^{m-1}\right)\bar{l}_{a}\left(a^{*m-1}\right)}\frac{q_{a}\left(a^{m-1}\mid a^{*m}\right)}{q_{a}\left(a^{*m}\mid a^{m-1}\right)}.
\]
Here $p_{a}$ is the prior distribution on $a$: as in \citet{constantine_accelerating_2016}
we use the factorisation $p_{a,i}\left(a,i\right)=p_{a}\left(a\right)p_{i\mid a}\left(i\mid a\right)$
of the prior on the active and inactive variables, and assume that
we can evaluate $p_{a}$ and $p_{i}$ pointwise, and that we can simulate
from $p_{i}\left(\cdot\mid a\right)$. This is the case, for example,
when a multivariate Gaussian prior is used on $\theta$. $\bar{l}_{a}(a)$
is an estimate of the marginal likelihood
\begin{equation}
l_{a}\left(a\right):=\int_{i}p_{i\mid a}\left(i\mid a\right)l\left(B_{a}a+B_{i}i\right)di.\label{eq:llhd}
\end{equation}
at point $a$. For $\bar{l}_{a}(a)$, \citet{constantine_accelerating_2016}
uses the Monte Carlo estimate
\begin{equation}
\bar{l}_{a}\left(a\right)=\frac{1}{N_{i}}\sum_{n=1}^{N_{i}}l\left(B_{a}a+B_{i}i^{n}\right)\label{eq:prior_llhd_estimator}
\end{equation}
where $i^{n}\sim p_{i\mid a}\left(\cdot\mid a\right)$ for $n=1:N_{i}$,
with $i^{n}$ being drawn each time an estimate $\bar{l}_{a}\left(a\right)$
is used.

\subsubsection{A pseudo-marginal approach\label{subsec:A-pseudo-marginal-approach}}

Although \citet{constantine_accelerating_2016} appears to succeed
in constructing an MH algorithm on the active variables $a$, there
are a number of reasons why this approach may be of limited use in
practice. As noted in \citet{constantine_accelerating_2016}, this
approach does not yield an MCMC algorithm with $\pi$ as its target
distribution: the algorithm is approximate. This point is discussed
futher in \citet{schuster_exact_2017} made the observation that this
approximate method can be made exact by formulating it as a pseudo-marginal
algorithm \citep{Beaumont2003,Andrieu2009} by only reestimating the
likelihood in the numerator at each iteration, i.e. by using the acceptance
probability
\[
\alpha_{a}^{m}=1\wedge\frac{p_{a}\left(a^{*m}\right)\bar{l}_{a}\left(a^{*m}\right)}{p_{a}\left(a^{m-1}\right)\bar{l}_{a}^{m-1}}\frac{q_{a}\left(a^{m-1}\mid a^{*m}\right)}{q_{a}\left(a^{*m}\mid a^{m-1}\right)},
\]
where $\bar{l}_{a}^{m-1}$ is the estimated marginal likelihood calculated
at the previous iteration of the MH. In a pseudo-marginal method,
any unbiased estimator may be used to estimate the marginal likelihood
IS. \citet{schuster_exact_2017} propose to use the importance sampling
(IS) estimator

\begin{equation}
\bar{l}_{a}\left(a\right)=\frac{1}{N_{i}}\sum_{n=1}^{N_{i}}\frac{p_{i\mid a}\left(i^{n}\mid a\right)l\left(B_{a}a+B_{i}i^{n}\right)}{q_{i}\left(i^{n}\mid a\right)},\label{eq:llhd_estimator}
\end{equation}
where $i^{n}\sim q_{i}\left(\cdot\mid a\right)$ for $n=1:N_{i}$.
We use the same notation $\bar{l}_{a}$ here, noting that the estimator
in equation (\ref{eq:prior_llhd_estimator}) is a special case of
the IS estimator, where the prior is used as a proposal. We refer
to the resultant method as active subspace MH (AS-MH). The full algorithm
is given in the Appendix.

This perspective not only `fixes' an issue with the \citet{constantine_accelerating_2016}
approach; it also brings into focus some of the properties of the
method. Firstly, a pseudo-marginal approach is justified \citep{Andrieu2009}
through its construction as a Metropolis-Hastings algorithm on the
joint space of $a$ and $\left\{ i^{n}\right\} _{n=1}^{N_{i}}$. Thus
it is potentially misleading to view the method as being a MH on only
the active variables $a$: the key point to consider is as to whether
this construction is likely to be more efficient than an MH on the
original parameterisation. From \citet{Pitt2012,Sherlock2015} it
is established that pseudo-marginal approaches can be efficient when
the variance of the log of the marginal likelihood estimator is of
the order of 1; empirical studies (e.g. \citet{gill2023bayesian,everitt2024ensemble})
suggest that when this variance is more than around 10 the MCMC mixes
poorly. In the following sections we discuss the implications of this.

\subsection{Tuning AS-MH}

The statistical efficiency AS-MH depends on the choice of $q_{i}$,
in conjunction with the choice of AS defined by $B_{a}$ and $B_{i}$.
The optimal choice of $q_{i}\left(i\mid a\right)$, in terms of minimising
the variance of equation (\ref{eq:llhd_estimator}) would be to choose
it proportional to $p_{i\mid a}\left(i\mid a\right)l\left(B_{a}a+B_{i}i\right)$.
The use of an AS allows us to use a proposal close to this optimal
choice, since the AS is chosen so that, in an ideal case, the inactive
variables are not at all influenced by the likelihood. In this case,
for any $a$, $l\left(B_{a}a+B_{i}i\right)\propto1$ as $i$ varies,
thus the optimal choice of $q_{i}\left(i\mid a\right)$ is the prior
$p_{i\mid a}\left(i\mid a\right)$. When this distribution is available
to us we can implement an `ideal' pseudo-marginal method. We can
see the AS as offering a criterion through which we may determine
when the pseudo-marginal approach, with the prior $p_{i\mid a}$ as
the proposal, might be an effective sampler.

Despite this appealing idea, in most cases the linear reparameteriation
used in an active subspace will not result in variables that are `perfectly'
inactive. For the remainder of this section we discuss the implications
of this.

\subsubsection{Choosing the active subspace\label{subsec:Choosing-the-active}}

To choose the AS, \citet{constantine_accelerating_2016,schuster_exact_2017}
use the Monte Carlo approximation in equation (\ref{eq:est_as}) to
estimate equation (\ref{eq:as}). They then identify a spectral gap
in the eigendecomposition to choose which directions are designated
as active and inactive variables: identifying a group of dominant
eigenvalues that are well separated from the remainder, and choosing
the corresponding directions to give the active variables. This procedure
is similar to the approach used in PCA for dimensionality reduction,
where the directions that explain a large proportion (e.g. 90\%) of
the variance are selected.

The framing of AS-MH as a pseudo-marginal approach suggests an alternative
means for selecting the inactive variables: one that is more directly
related to the efficiency of the resulting MCMC. Recall that for a
pseudo-marginal method to be effective the estimator in equation (\ref{eq:llhd_estimator})
should have a low variance. The \emph{effective sample size} (ESS)
\citep{Kong1994}
\begin{equation}
\mbox{ESS}=\left(\sum_{n=1}^{N}\left(w^{n}\right)^{2}\right)^{-1}.\label{eq:ess}
\end{equation}
of an IS with weights $\left\{ w^{n}\right\} _{n=1}^{N}$ gives an
estimate of the effective number of points of the IS estimator (compared
to the standard Monte Carlo estimator). A new approach to selecting
the inactive variables would be to select the largest number of inactive
variables (starting with those than have the smallest eigenvalues)
that yields an ESS that does not drop below some threshold or, for
reasons described later in the paper, yields a variance of the log-likelihood
that is not larger than some threshold (of order 1) for some value
in the typical set of the posterior of $a$.

\subsubsection{Implications for efficiency of AS-MH}

The schemes of \citet{constantine_accelerating_2016,schuster_exact_2017}
are unlikely to be effective in many situations, for the following
reason. The situation where active subspaces are most likely to be
useful in improving the efficiency of an MCMC algorithm is the case
when the procedure in section \ref{subsec:Choosing-the-active} identifies
a large number of inactive variables. In this case, the dimension
of the pseudo-marginalised space is greatly reduced and, as long as
the prior is a good proposal for the inactive variables, the MCMC
efficiency will be increased. However, the prior is only an effective
proposal for the inactive variables for which the likelihood is exactly
constant. This condition does hold for some models (see e.g. \citet{drton_bayesian_2017}),
but it is much more common to find that most eigenvalues of equation
(\ref{eq:as}) are not precisely zero \citep{constantine_accelerating_2016,brouwer_underlying_2018},
so that the likelihood has some dependence on the inactive variable
(before even considering that in practice we work with the Monte Carlo
approximation in equation (\ref{eq:est_as})). In this situation,
we know that the number of points required to stablise the variance
of the IS estimator, and therefore the efficiency of the pseudo-marginal
MCMC algorithm, scales exponentially in the dimension of the inactive
variables. Since we desire to gain maximum benefit in high-dimensional
settings, this limits the usefulness of the algorithms in \citet{constantine_accelerating_2016,schuster_exact_2017}.

\subsection{Paper structure}

In this paper we propose to instead use particle marginal MH (PMMH)
\citep{Andrieu2010c} to make use of active spaces in Monte Carlo
methods without encountering the poor performance indicated by the
argument above. The main idea is to replace the IS estimator for active
variable marginal likelihood proposed by \citet{schuster_exact_2017}
with an SMC approach, which scales better the dimension of the inactive
variables \citep{Beskos2014b}. We investigate empirically these approaches,
and use these findings to propose alternatives in section \ref{sec:Metropolis-within-Gibbs-and-part},
before a discussion in section \ref{sec:Conclusions}.

\section{Particle marginal MH\label{sec:Particle-marginal-MH}}

\subsection{SMC sampler recap\label{subsec:SMC-sampler-recap}}

An SMC sampler \citep{DelMoral2006c} simulates from a target distribution
$\pi_{T}$ through propagating a population of weighted importance
points (known as `particles') through a sequence of distribtuions
$\pi_{0},...,\pi_{T}$. In this sequence it is possible to simulate
from $\pi_{0}$, then a sequential IS approach is used to successively
draw points from each target, using: (typically) MCMC moves to move
the points; IS reweighting to account for the discrepancy between
the distributions the points are drawn from and the desired target;
and resampling steps to avoid particles with small weights being propagated.
Throughout this paper we use stratified resampling \citep{douc_comparison_2005}.

We let $\tilde{\pi}_{t}$ denote an unnormalised version of $\pi_{t}$
for each $t$. The choice of the sequence of distributions significantly
affect the performance of the algorithm. We consider the case where
$\tilde{\pi}_{T}(\theta)=p(\theta)l(\theta)$, where $p$ is a prior
and
\[
l\left(\theta\right)=\prod_{s=1}^{T}l_{s}\left(\theta\right),
\]
for some sequence $l_{s}\left(\theta\right)$. We use the notation
\[
l_{1:t}\left(\theta\right)=\prod_{s=1}^{t}l_{s}\left(\theta\right),
\]
so that $l_{1:T}\left(\theta\right)=l\left(\theta\right)$. The SMC
samplers in this paper use $\tilde{\pi}_{t}(\theta)=p(\theta)l_{1:t}\left(\theta\right)$.
This encompasses the data point tempering of \citet{Chopin2002} (where
the factors each depend on a different data point), and also the widely-used
annealing approach (where $l\left(\theta\right)=\prod_{s=1}^{T}l^{\eta_{t}-\eta_{t-1}}\left(\theta\right)$
for $\eta_{0}=0<...<\eta_{T}=1$).

\subsection{PMMH with active subspaces\label{sec:Particle-MCMC-with}}

To construct a PMMH extension to the AS-MH algorithm we follow precisely
the framework in \citet{Andrieu2010c}. PMMH in our context boils
down to running an MH algorithm on the active variables, where the
marginal likelihood in equation (\ref{eq:llhd}) at the proposed point
at each iteration is estimated by an SMC sampelr, rather than the
importance sampler in AS-MH. We call the resultant algorithm active
subspace PMMH (AS-PMMH). The sequence of targets used in this SMC
sampler is $\pi_{0,i}\left(i\mid a\right)=p_{i\mid a}\left(i\mid a\right)$
and, for $t=1:T$,
\[
\pi_{t,i}\left(i\mid a\right)=\frac{p_{i\mid a}\left(i\mid a\right)l_{1:t}\left(B_{a}a+B_{i}i\right)}{l_{t,a}\left(a\right)},
\]
where $l_{t,a}\left(a\right)$ is the normalising constant for $\tilde{\pi}_{t,i}\left(i\mid a\right)=p_{i\mid a}\left(i\mid a\right)l\left(B_{a}a+B_{i}i\right)$.
We use the SMC sampler to estimate this normalising constant at iteration
$T$, which is the marginal likelihood of the active variables $l_{a}(a)$
from equation (\ref{eq:llhd}), with
\[
\hat{l}_{T,a}\left(a\right)=\prod_{t=1}^{T}\sum_{n=1}^{N_{i}}\tilde{w}_{t}^{n},
\]
where $\tilde{w}_{t}^{n}$ is the unnormalised weight in the SMC sampler
of particle $n$ at iteration $t$. The SMC sampler is given in full
in algorithm \ref{alg:inactive_smc}, with AS-PMMH being given in
algorithm \ref{alg:AS-PMMH}. We use the output of algorithm \ref{alg:AS-PMMH}
in the same way as AS-MH, using estimators (\ref{eq:est1}) and (\ref{eq:est2}).

\begin{algorithm} \caption{SMC on inactive variables for a given $a$ and $t$.}\label{alg:inactive_smc}

Simulate $N_{i}$ points, $\left\{ i_{0}^{n} \right\}_{n=1}^{N_{i}} \sim p_{i\mid a}\left( \cdot \mid a \right)$ and set each weight $w_0^n = 1/N_i$;\

\For {$s=1:t$}
{
	\For(\tcp*[h]{reweight}) {$n=1:N_i$}
	{
		\eIf {$s=1$}
		{
			\[
			\tilde{w}^n_{s} = w^n_{s-1} l_{1:s}\left(B_{a} a+B_{i} i^{n}\right);
			\]
		}
		{
			\[
			\tilde{w}^n_{s} = w^n_{s-1} \frac{l_{1:s}\left(B_{a} a+B_{i} i_{s-1}^{n}\right)}{l_{1:s-1}\left(B_{a}a+B_{i}i_{s-1}^{n}\right)};
			\]
		}
	}

	$\left\{ w^n_{s} \right\}_{n=1}^{N_{i}} \leftarrow \mbox{ normalise}\left( \left\{ \tilde{w}^n_{s} \right\}_{n=1}^{N_{i}} \right)$;

	If $s=t$, go to line 32;

	\For{$n=1:N_i$}
	{
		Simulate the index $v^n_{s-1} \sim \mathcal{M}\left( \left( w_s^1, ..., w_s^{N_i} \right) \right)$ of the ancestor of particle $n$;
	}

	\If(\tcp*[h]{resample}) {some degeneracy condition is met}
	{
		\For{$n=1:N_i$}
		{
			Set $i^{n}_{s} = i^{v^n_{s-1}}_{s-1}$;
		}
		$w^n_{s} = 1/N_i$ for $n=1:N_i$;
	}
	\Else
	{
		\For{$n=1:N_i$}
		{
			Set $i^{n}_{s} = i^{n}_{s-1}$;
		}
	}
	
	\For(\tcp*[h]{move}) {$n=1:N_i$}
	{
		Simulate $i^{n*}_{s} \sim q_{t,i} \left( \cdot \mid i^{n}_{s}, a \right)$;
		
		Set $i^{n}_{s} = i^{n*}_{s}$ with probability

		\[
		1\wedge\frac{l_{1:s}\left(B_{a}a+B_{i}i^{n*}_{s}\right)p_{i\mid a}\left( i^{n*}_{s} \mid a \right) q_{t,i} \left( i^{n}_{s} \mid a \right)}{l_{1:s}\left(B_{a}a+B_{i}i^{n}_{s}\right)p_{i\mid a}\left( i^{n}_{s} \mid a \right) q_{t,i} \left( i^{n*}_{s} \mid a \right)},
		\]
	}
}
Estimate $l_{t,a}\left( a \right)$ using
\[
\hat{l}_{t,a}\left(a\right) = \prod_{s=1}^t \sum_{n=1}^{N_{i}} \tilde{w}^n_{s}.
\]
\end{algorithm}

\begin{algorithm}
\caption{Active subspace particle marginal Metropolis-Hastings}\label{alg:AS-PMMH}

Initialise $a^0$;\

Run algorithm \ref{alg:inactive_smc} for $a=a^0$ and $t=T$, obtaining $\hat{l}_{T,a}\left(a^{0} \right)$ and weights, denoted $\left( w_T^{1,0}, ..., w_T^{N_i,0} \right)$;\

$u^{0} \sim\mathcal{M}\left( \left( w_T^{1,0}, ..., w_T^{N_i,0} \right) \right)$;\

Let $\hat{l}^0_{a} = \hat{l}_{T,a}\left(a^{0} \right)$;\

\For {$m=1:N_a$}
{
	$a^{*m} \sim q_a\left(\cdot\mid a^{m-1} \right)$;\

	Run algorithm \ref{alg:inactive_smc} for $a=a^{*m}$ and $t=T$, obtaining $\hat{l}_{T,a}\left(a^{*m} \right)$ and  weights, denoted $\left( w_T^{*1,m}, ..., w_T^{*N_i,m} \right)$;\

	$u^{*m} \sim\mathcal{M}\left( \left( w_T^{*1,m}, ..., w_T^{*N_i,m} \right) \right)$;\

    Set $\left(a^m, \left\{ i^{n,m}, w^{n,m} \right\}_{n=1}^{N_i}, u^m, \hat{l}^m_{T,a} \right) = \left(a^{*m},\left\{ i^{*n,m}, w^{*n,m} \right\}_{n=1}^{N_i}, u^{*m}, \hat{l}_{T,a}\left(a^{*m} \right) \right)$ with probability
	\[
	\alpha_a^m = 1\wedge\frac{p_{a}\left(a^{*m} \right) \hat{l}_{T,a}\left(a^{*m} \right)}{p_{a}\left(a^{m-1} \right) \hat{l}^{m-1}_{T,a}}\frac{q_a\left(a^{m-1} \mid a^{*m}\right)}{q_a\left(a^{*m}\mid a^{m-1} \right)};
	\]\
	
	Else let $\left(a^m, \left\{ i^{n,m}, w^{n,m} \right\}_{n=1}^{N_i}, u^m, \hat{l}^m_{T,a} \right) = \left(a^{m-1},\left\{ i^{n,m-1}, w^{n,m-1} \right\}_{n=1}^{N_i}, u^{m-1}, \hat{l}^{m-1}_{T,a}\right)$;\

}
\end{algorithm}

\subsection{Empirical investigation\label{subsec:Empirical-investigation}}

\subsubsection{Model\label{subsec:Model}}

\begin{landscape}

\begin{figure}
\subfloat[{A 2d slice of the Gaussian prior on the horizontal plane $\theta_{1}=0$
together with the level surface of the likelihood in the particular
case $\sum_{k=1}^3 \theta_{k}=0$ (green plane).\label{fig:A-2d-slice}}]{\includegraphics[scale=0.3]{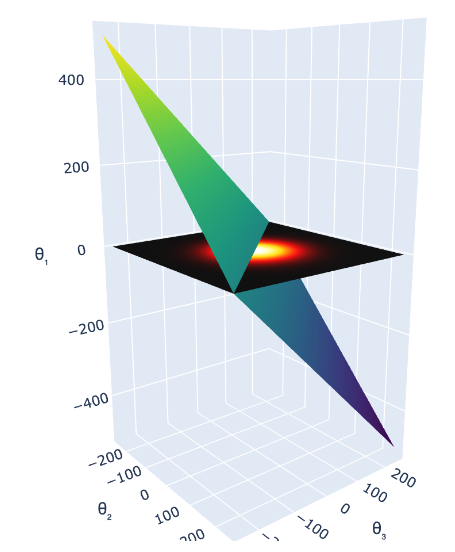}

}\subfloat[{The eigenvalues used in determining the active subspace.\label{fig:The-eigenvalues-used}}]{\includegraphics[scale=0.35]{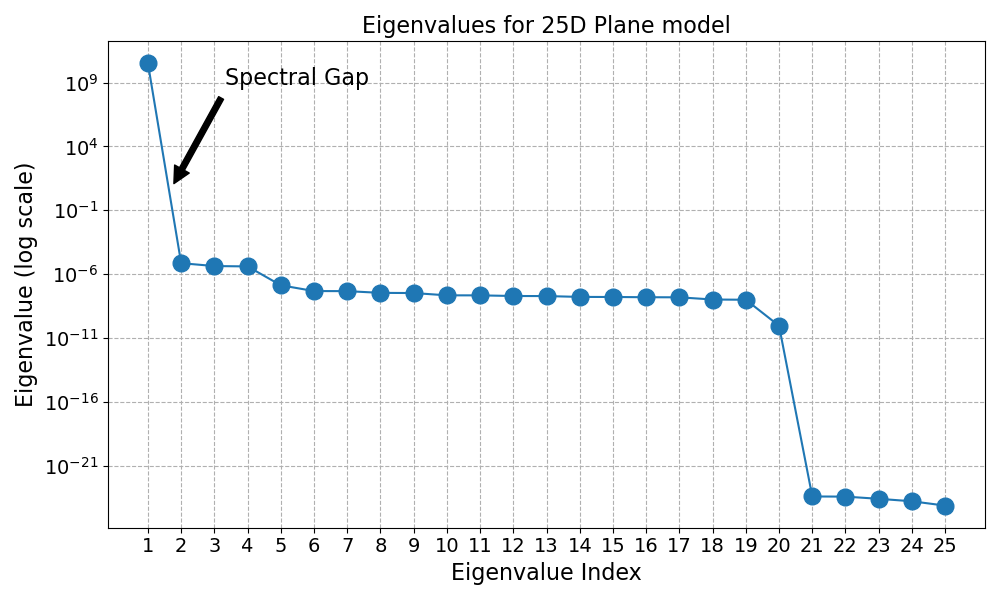}

}\subfloat[{The ESS of the IS marginal likelihood estimator against the active
subspace dimension for the plane model.\label{fig:The-ESS-of}}]{\includegraphics[scale=0.35]{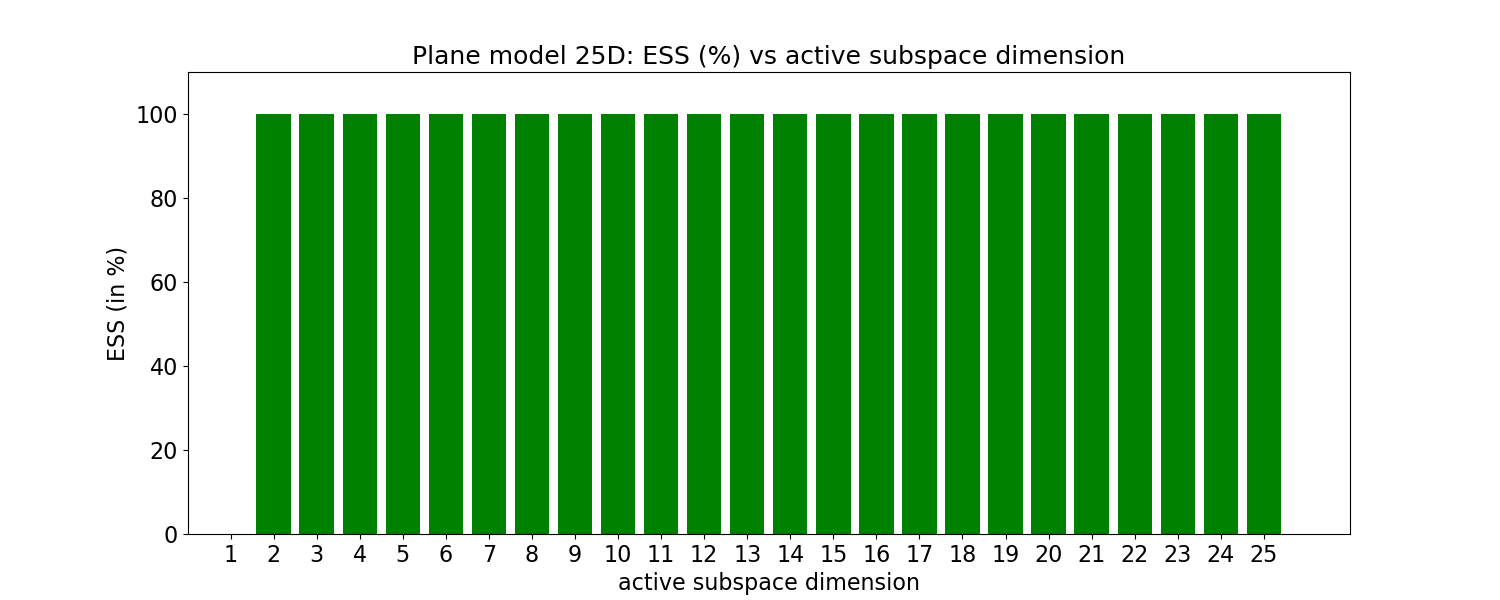}

}

\subfloat[{A 2d slice of the Gaussian prior on the horizontal plane $\theta_{1}=0$
together with the level surface of the likelihood in the particular
case $\sum_{k=1}^3 \theta_{k} + 0.001 \times \sum_{k=1}^3 \theta^2_{k}=0$ (green).\label{fig:A-2d-slice-1}}]{\includegraphics[scale=0.4]{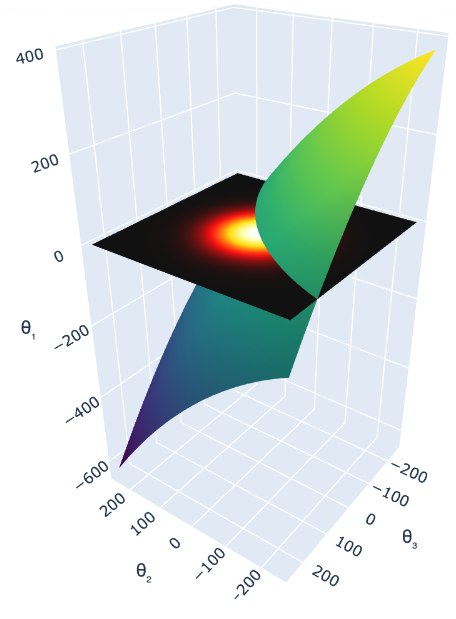}

}\subfloat[{The eigenvalues used in determining the active subspace.\label{fig:The-eigenvalues-used-1}}]{\includegraphics[scale=0.35]{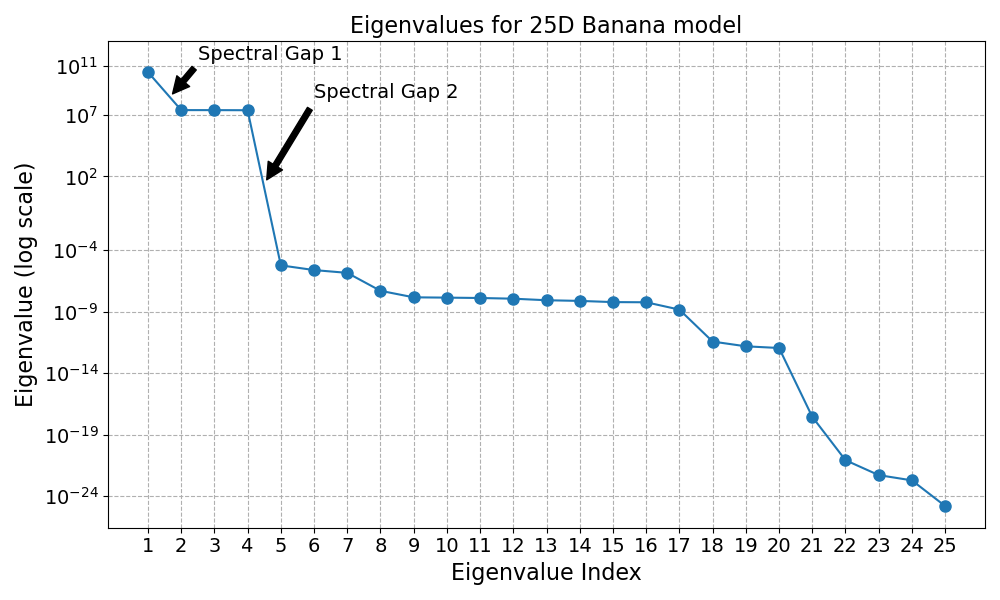}

}\subfloat[{The ESS of the IS marginal likelihood estimator against the active
subspace dimension for the banana model.\label{fig:The-ESS-of-1}}]{\includegraphics[scale=0.35]{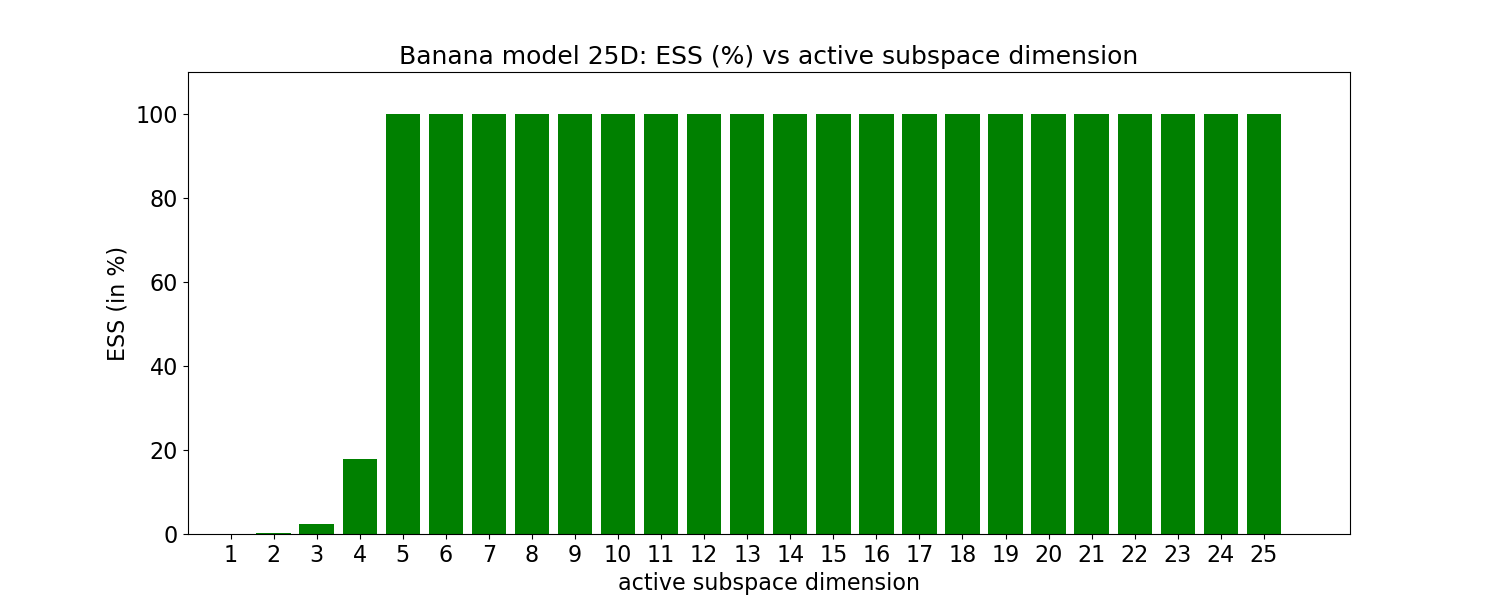}

}

\caption{{The plane and banana models: illustration and results.}}
\end{figure}

\end{landscape}

As an example of where active subspaces are applicable, we use a toy
model where we artificially ensure that large numbers of variables
are not involved in the likelihood. We let $\theta=(\theta_{1},...,\theta_{d})^{T}\in\mathbb{R}^{d}$,
and perform inference for this parameter given observations of variable
$y\in\mathbb{R}$. Let $\theta_{i}\sim\mathcal{N}\left(0,\sigma=\sqrt{5000}\right)$
be the prior distribution and let $y\sim\mathcal{N}\left(\sum_{i=1}^{d}\theta_{i},1\right)$
be the model for observed data. We use as data 100 observations of
$y$ drawn from $\mathcal{N}(0,1)$. We have that the $\theta_{i}$
can only be identified as lying on a hyperplane of dimension $d-1$
with equation $\sum_{i=1}^{d}\theta_{i}=0$. Figure \ref{fig:A-2d-slice}
gives an illustration of this model for $d=3$. We call this model
the `plane' model.

This model is tailor-made for active subspace approaches, but this
type of structure is extremely unlikely to be encountered in a realistic
model. The key feature of the model is that there is a ridge in the
posterior, as is found in some posterior distributions with identifiability
issues. However, this ridge is linear: a feature that is very unrealistic.
We consider instead the model $y\sim\mathcal{N}\left(\sum_{i=1}^{d}\theta_{i}+b\sum_{j=1}^{k}\theta_{i}^{2},1\right)$,
for $1\leq k\le d$, which will result in a curve in the `plane'.
Figure \ref{fig:A-2d-slice-1} gives an illustration of this model
for $d=3$, $k=2$ and $b=0.001$. We call this model the `banana'
model, after the shape of the posterior it produces in two dimensions.

Figures \ref{fig:The-eigenvalues-used} and \ref{fig:The-eigenvalues-used-1}
show, respectively, the eigenvalues in the decomposition used to identify
the active subspace in the plane (parameters $d=25$) and banana (parameters
$d=25$, $k=3$) models. We see in the plane model that a spectral
gap can be clearly identified between the 1st and 2nd eigenvalue,
with all but the first eigenvalues being very small (there is another
gap, between the 19th and 20th eigenvalues, but at this point the
eigenvalues are so small this gap is not important). Here we might
expect a large benefit from using an AS of dimension 1. In the banana
model, which has a similar structure in some respects, we see a gap
between the first and second eigenvalues, then another gap between
the fourth and fifth eigenvalues. Following the reasoning in \citet{constantine_accelerating_2016},
one might identify two possible active subspaces: one of one dimension,
where the inactive variables lie on the curved `plane'; and another
of four dimensions where the variables affected by the curvature all
need to be added to the AS.

We study this further using the ESS approach \ref{subsec:Choosing-the-active}
to selecting the AS. We estimated the ESS for the IS estimator of
the marginal likelihood at $a=\mathbf{0}_{d_{a}}$: the ESS as a percentage
of the $N_{i}=10^{4}$ inactive points is plotted against the dimension
of the inactive subspace, for the plane and banana models in figures
and respectively. For the plane model, we deduce that the estimation
of the marginal likelihood will have a small error when using up to
24 inactive variables, indicating a one-dimensional AS. For the banana
model, we see that the ESS indicates a choice of AS of dimension 4.
The alternative choice of dimension 1 indicated in the eigenvalue
plot is shown to be ineffective, since this choice would result in
a marginal likelihood estimate with a low ESS (and hence a high variance)
and an ineffiicient AS-MH algorithm. From this analysis we draw two
conclusions. Firstly, we recommend the use of the ESS diagnostic in
choosing the AS, since it is more indicative of the likely performance
of an AS-MH algorithm resulting from a particular choice of AS. Secondly,
the weaknesses of the linearity assumption behind active subspaces
is highlighted: when a slight curve is introduced into the plane model
(giving the banana model), all of the variables affected by the curvature
are required in the AS.

\subsubsection{Performance of AS-MH and AS-PMMH\label{subsec:Performance-of-AS-MH}}

We examine the performance of MH, AS-MH and AS-PMMH on the plane and
the banana models, with active subspace dimension 1: so that active
dimension is perpendicular to the plane (or curved plane in the banana
model). Each algorithm used $10^{5}$ likelihood evaluations: MH used
$10^{5}$ MCMC iterations; AS-MH used $N_{i}=10$ inactive points
and $10^{4}$ MCMC iterations; AS-PMMH used $N_{i}=10$ inactive points,
6 intermediate distributions in the SMC used in the marginal likelihood
estimates, and 1666 MCMC iterations. The proposal on the active variables
(or $\theta$ for MH) was chosen to be $2.38^{2}/d_{a}$ times the
estimated posterior covariance from a pilot MCMC run, this being an
approximation to the optimal proposal in the case of a Gaussian target
and proposal. The proposal for the MCMC move on the inactive variables
was chosen adaptively as the algorithm ran, using at each iteration
$2.38^{2}/d_{i}$ times the estimated covariance of the points from
the previous SMC iteration.

\begin{figure}
\subfloat[{The error of estimates of the posterior mean on the 25d plane model.\label{fig:The-error-of}}]{\includegraphics[scale=0.12]{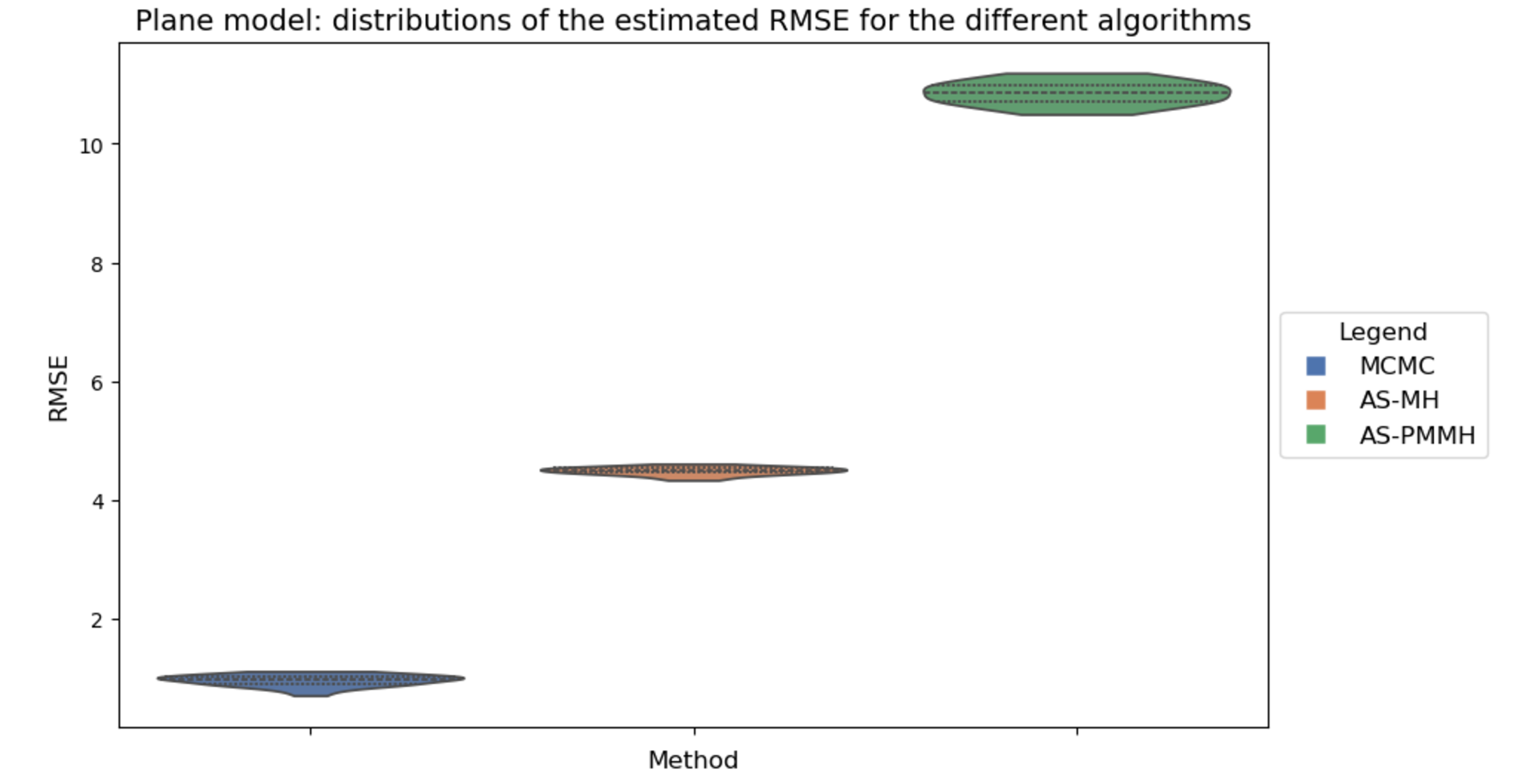}

}\subfloat[{The error of estimates of the posterior mean on the 25d banana model.\label{fig:The-error-of-1}}]{\includegraphics[scale=0.34]{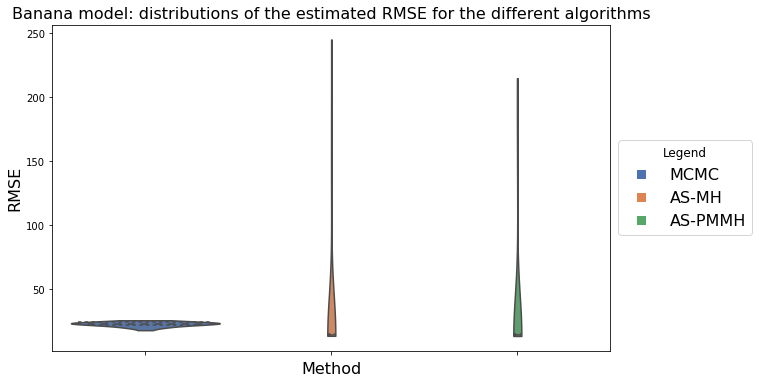}

}\caption{{Comparing MH, AS-MH and AS-PMMH.}}
\end{figure}

The errors of posterior mean estimates on the 25d plane and banana
models are shown in figures \ref{fig:The-error-of} and \ref{fig:The-error-of-1}
respectively. We see that the use of an AS in AS-MH and AS-PMMH does
not compensate for the smaller number of iterations used in the MCMC,
even in the plane model which has been designed to be favourable to
AS-MH. In the banana model we observe the effect of the characteristically
`sticky' behaviour of pseudo-marginal algorithms.

One might consider parallel implementations, drawing the inactive
variables in parallel, that might result in AS-MH and AS-PMMH being
competative with MH. However, in such cases, to achieve a reduction
in cost for the AS methods we would need the overheads associated
with parallelisation to be negligible compared to the cost of the
likelihood evaluation. Our conclusion is therefore that AS-MH and
AS-PMMH are unlikely to be of practical use in most models. The main
drawback is that much of our computational effort is devoted to simulating
the inactive variables, which should be the `easiest' to simulate.
In the following section we propose new AS methods to remedy this
issue.

\section{Metropolis-within-Gibbs and particle Gibbs\label{sec:Metropolis-within-Gibbs-and-part}}

Section \ref{subsec:Performance-of-AS-MH} illustrates the poor performance
of AS-MH and AS-PMMH. In this section we propose new approaches that
make use of AS, but avoid the two pitfalls of AS-MH and AS-PMMH:
\begin{enumerate}
\item The use of multiple inactive points per active point, which focusses
more computational effort on the simulation of inactive variables
than active variables.
\item The use of the pseudo-marginal approach, which can be very inefficient
if the variance of the marginal likelihood estimates is high.
\end{enumerate}
These requirements lead us to a surprisingly simple and effective
idea: the use of a Metropolis-within-Gibbs approach. We detail this
method in section \ref{subsec:Metropolis-within-Gibbs}, with an extension
to using SMC approaches on the active variables in section \ref{subsec:Metropolis-within-particle-Gibbs}.

\subsection{Metropolis-within-Gibbs\label{subsec:Metropolis-within-Gibbs} }

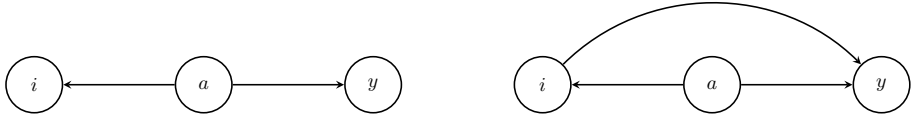
\begin{figure}[htb] \centering \resizebox{12cm}{!}{\begin{tikzpicture}[
        node distance = 3cm,
        thick,
        node/.style = {circle, draw, minimum size=1cm},
        arrow/.style = {->, >=stealth}
    ]
    
\node[node] (i1) at (0,0) {$i$};
\node[node] (a1) [right of=i1] {$a$};
\node[node] (y1) [right of=a1] {$y$};

\draw[arrow] (a1) -- (i1);
\draw[arrow] (a1) -- (y1);

\node[node] (i2) at (9,0) {$i$};
\node[node] (a2) [right of=i2] {$a$};
\node[node] (y2) [right of=a2] {$y$};

\draw[arrow] (a2) -- (i2);
\draw[arrow] (a2) -- (y2);
\draw[arrow] (i2) to[out=45, in=135, looseness=1] (y2);
    
    \end{tikzpicture}
 } \caption{A graphical model illustrating the case of perfect inactive variables (left) and the general case (right).}\label{fig:gm} \end{figure}

In the ideal AS case we choose inactive variables so that they are
not influenced at all by the observed data, except through the prior
$p_{a,i}$; this parameterisation is illustrated on the left of figure
\ref{fig:gm}. This situation would suggest the following MCMC `Gibbs'-style
approach for drawing points from the posterior of $a$ and $i$: simulate
$i\sim\pi_{i\mid a}(i\mid y,a)\propto p_{i\mid a}(i\mid a)$; simulate
$a\sim\pi_{a\mid i}(a\mid y,i)\propto p_{a}(a)f(y\mid a)$. The efficiency
of this approach depends heavily on the degree of posterior dependence
of $a$ and $i$. Due to the structure of the model, the posterior
dependence is here precisely the same as the prior dependence. There
are many situations in which we might expect this prior dependence
to be small: in particular, if a multivariate Gaussian prior with
diagonal covariance is chosen for $\theta$, $a$ and $i$ would be
\emph{a priori} (and hence \emph{a posteriori}) independent. Therefore
we might expect the Gibbs approach to be efficient. If the draw from
$\pi_{a\mid i}$ cannot be made exactly, a Metropolis-Hastings move
may be used in its place.

Outside of the ideal case, we instead have the dependence structure
on the right of figure \ref{fig:gm}. This structure changes the full
conditional distibution $\pi_{i\mid a}$ used in the first step of
the Gibbs approach. We now have $i\sim\pi_{i\mid a}(i\mid y,a)\propto p_{i\mid a}(i\mid a)f(y\mid a)$,
and in general need to used a Metropolis-Hastings move in place of
exact simulation here. Nevertheless, if the inactive variables only
have a weak posterior dependence on the active variables, we might
still expect this approach to be efficient. We call this algorithm,
shown in full in algorithm \ref{alg:asmwg}, active subspace Metropolis-within-Gibbs
(AS-MwG).

\begin{algorithm}
\caption{Active subspace Metropolis within Gibbs}\label{alg:asmwg}

Initialise $i^0$ and $a^{0}$;\

\For {$m=1:N$}
{
	$i^{*m} \sim q_i\left(\cdot\mid i^{m-1},a^{m-1} \right)$;\

	Set $i^m = i^{*m}$ with probability
	\[
	\alpha_i^m = 1\wedge \frac{p_{i\mid a}\left(i^{*m} \mid a^{m-1}\right)l\left(B_{i}i^{*m}+B_{a}a^{m-1}\right) q_i\left(i^{m-1} \mid i^{*m}, a^{m-1} \right)}{p_{i \mid a}\left(i^{m-1} \mid a^{m-1}\right)l\left(B_{i}i^{m-1}+B_{a}a^{m-1}\right) q_i\left(i^{*m} \mid i^{m-1}, a^{m-1} \right)};
	\]\
	
	Else let $i^m=i^{m-1}$;\

	$a^{*m} \sim q_a\left(\cdot\mid a^{m-1}, i^{m} \right)$;\

	Set $a^m = a^{*m}$ with probability
	\[
	\alpha_a^m = 1\wedge \frac{p_{a}\left(a^{*m}\right)l\left(B_{i}i^{m}+B_{a}a^{*m}\right) q_a\left(a^{m-1} \mid a^{*m}, i^m \right)}{p_{a}\left(a^{m-1} \right) l\left(B_{i}i^{m}+B_{a}a^{m-1}\right) q_a\left(a^{*m} \mid a^{m-1}, i^m \right)};
	\]\
	
	Else let $a^m=a^{m-1}$;\
}
\end{algorithm}

AS-MwG requires the choice of two proposals: $q_{i}$ and $q_{a}$.
Since we expect the full conditional $\pi_{i\mid a}$ to be close
to $p_{i\mid a}$, we choose $q_{i}\left(i^{*m}\mid i^{m-1},a^{m-1}\right)=p_{i\mid a}\left(i^{*m}\mid a^{m-1}\right)$.
In the ideal case, when $f$ is constant over different choices of
$i$, this results in an acceptance probability $\alpha_{a}^{m}$
of 1. For $q_{a}$, we may use any standard approach for tuning the
proposal: this is simplified in the ideal case since we need not condition
the proposal on $i$.

This method maintains the advantages of AS-MH, in that it makes use
of the AS to improve the design of the MCMC, but avoids the drawback
of needing to estimate the marginal likelihood of $a$. For AS-MH,
the estimation of the marginal likelihood with IS becomes very inefficient
outside of the ideal AS case. We will examine the performance of AS-MwG
in such a situation in section \ref{sec:Empirical-results}.

\subsection{Metropolis-within-particle Gibbs\label{subsec:Metropolis-within-particle-Gibbs}}

We may extend the AS-MwG approach such that SMC may be used to simulate
the active variables. This may be of use when it is difficult to design
an effective proposal to simulated the active variables (since we
could use adaptive SMC techniques), or when their simulation would
benefit from a population approach (for example when there are multiple
modes).

The construction of this algorithm directly follows from the particle
MCMC framework. Following the nomenclature in \citet{Andrieu2010c},
this method is the particle Gibbs counterpart to the PMMH algorithm
introduced in section \ref{sec:Particle-MCMC-with}, were we to swap
the roles of the active and inactive variables. The reason for swapping
the roles is the observation in \ref{subsec:Performance-of-AS-MH}
that a drawback of AS-MH and AS-PMMH is that they spend more computational
effort on the inactive variables than the inactive. We reverse these
roles so that more effort is focussed on the active variables. We
note that, as is backed up by empirical results below, this is unlikely
to be a productive strategy for a PMMH algorithm: it would require
the estimation of the marginal likelihood of the inactive variables
(using SMC on the active variables), which due to the active variables
being `difficult' to simulate. is likely to have a high variance.

In algorithm \ref{alg:asmwpg} we give full details for this new method,
which we call AS Metropolis within particle Gibbs (AS-MwPG). This
algorithm calls a conditional SMC algorithm on the active variables,
given in algorithm \ref{alg:active_csmc}. Conditional SMC is a standard
SMC algorithm, except that one particle trajactory is given as input
to the algorithm, and this particle remains fixed throughout the algorithm.
\citet{Andrieu2010c} explains how this algorithm is the result of
using a particular MCMC move applied to the (extended) target distribution
used in particle MCMC methods.

\begin{algorithm}
\caption{Active subspace Metropolis within particle Gibbs}\label{alg:asmwpg}

Initialise $i^0$;\

Initialise $a^{1,0}_t$ for $t=0:T$;\

\For {$m=1:N_i$}
{
	$i^{*m} \sim q_i\left(\cdot\mid i^{m-1}, a_{T}^{1,m-1} \right)$;\

	Set $i^m = i^{*m}$ with probability
	\[
	\alpha_i^m = 1\wedge \frac{p_{i\mid a}\left(i^{*m} \mid a_{T}^{1,m-1}\right)l_{1:T}\left(B_{i}i^{*m}+B_{a}a_{T}^{1,m-1}\right) q_i\left(i^{m-1} \mid i^{*m}, a_{T}^{1,m-1} \right)}{p_{i\mid a}\left(i^{m-1} \mid a_{T}^{1,m-1}\right)l_{1:T}\left(B_{i}i^{m-1}+B_{a}a_{T}^{1,m-1}\right) q_i\left(i^{*m} \mid i^{m-1}, a_{T}^{1,m-1} \right)};
	\]\
	
	Else let $i^m=i^{m-1}$;\

	Run algorithm \ref{alg:active_csmc} for $i=i^m$, $a^1_{0:T} = a^{1,m-1}_{0:T}$, $\tilde{w}^{1,m-1}_{0:T}$  and $t=T$, obtaining points $\left( a_{0:T}^{1,m}, ..., a_{0:T}^{N_a,m} \right)$ and unnormalised and normalised weights $\left( \tilde{w}_T^{1,m}, ..., \tilde{w}_T^{N_a,m} \right)$ and $\left( w_T^{1,m}, ..., w_T^{N_a,m} \right)$;\

	$u^{m} \sim\mathcal{M}\left( \left( w_T^{1,m}, ..., w_T^{N_a,m} \right) \right)$;\

	Set $a_{0:T}^{1,m} = a_{0:T}^{u^{m},m}$;\
	
	Set $\tilde{w}_{0:T}^{1,m} = \tilde{w}_{0:T}^{u^{m},m}$;\

	Set $w_{0:T}^{1,m} = w_{0:T}^{u^{m},m}$;\
}
\end{algorithm}

\begin{algorithm} \caption{Conditional SMC on active variables for a given $i$, $a^1_{0:T}$, $\tilde{w}^1_{0:T}$ and $t$.}\label{alg:active_csmc}

Simulate $N_{a}-1$ points, $\left\{ a_{0}^{n} \right\}_{n=2}^{N_{a}} \sim p_a$ and set each weight $w_0^n = 1/N_a$;\

\For {$s=1:t$}
{
	\For(\tcp*[h]{reweight}) {$n=2:N_a$}
	{
		\eIf {$s=1$}
		{
			\[
			\tilde{w}^n_{s} = w^n_{s-1} l_{1:s}\left(B_{i} i+B_{a} a^{n}\right);
			\]
		}
		{
			\[
			\tilde{w}^n_{s} = w^n_{s-1} \frac{l_{1:s}\left(B_{i} i+B_{a} a_{s-1}^{n}\right)}{l_{1:s-1}\left(B_{i}i+B_{a}a_{s-1}^{n}\right)};
			\]
		}
	}

	$\left\{ w^n_{s} \right\}_{n=1}^{N_{a}} \leftarrow \mbox{ normalise}\left( \left\{ \tilde{w}^n_{s} \right\}_{n=1}^{N_{a}} \right)$;

	If $s=t$, terminate the algorithm;

	\For{$n=2:N_a$}
	{
		Simulate the index $v^n_{s-1} \sim \mathcal{M}\left( \left( w_s^1, ..., w_s^{N_a} \right) \right)$ of the ancestor of particle $n$;
	}

	\If(\tcp*[h]{resample}) {some degeneracy condition is met}
	{
		\For{$n=2:N_a$}
		{
			Set $a^{n}_{s} = a^{v^n_{s-1}}_{s-1}$;
		}
		$w^n_{s} = 1/N_a$ for $n=1:N_a$;
	}
	\Else
	{
		\For{$n=2:N_a$}
		{
			Set $a^{n}_{s} = a^{n}_{s-1}$;
		}
	}
	
	\For(\tcp*[h]{move}) {$n=2:N_a$}
	{
		Simulate $a^{n*}_{s} \sim q_{t,a} \left( \cdot \mid a^{n}_{s}, i \right)$;
		
		Set $a^{n}_{s} = a^{n*}_{s}$ with probability

		\[
		1\wedge\frac{l_{1:s}\left(B_{i}i+B_{a}a^{n*}_{s}\right)p_a\left( a^{n*}_{s} \right) q_{t,a} \left( a^{n}_{s} \mid i \right)}{l_{1:s}\left(B_{i}i+B_{a}a^{n}_{s}\right)p_a\left( a^{n}_{s} \right) q_{t,a} \left( a^{n*}_{s} \mid i \right)},
		\]
	}
}

\end{algorithm}

\subsection{Empirical results\label{sec:Empirical-results}}

\subsubsection{Metropolis-within-Gibbs}

\begin{figure}
\includegraphics[scale=0.8]{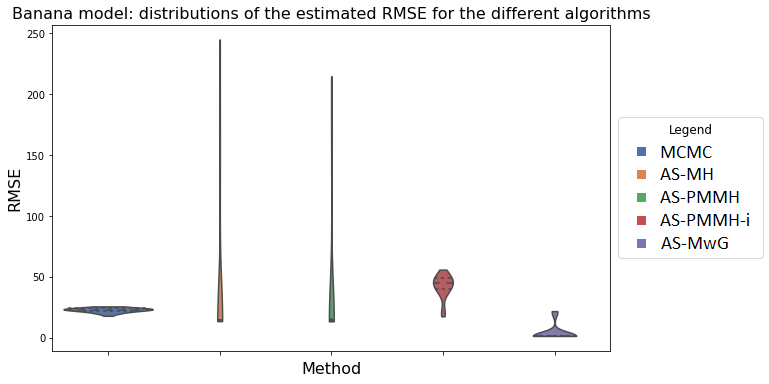}

\caption{Comparing AS-MwG with other methods on the banana model.\label{fig:Comparing-AS-MwG-with}}

\end{figure}

We applied AS-MwG to the banana model from section \ref{subsec:Model},
comparing it to MH, AS-MH, AS-PMMH and AS-PMMH with the role of the
active and inactive variables inverted (we name this AS-PMMH-i). We
used the model with $d=25$ and $k=3$, using an active subspace of
size 1, as in section \ref{subsec:Performance-of-AS-MH}. All algorithms
were run using $10^{5}$ likelihood iterations: MH used $10^{5}$
MCMC iterations; AS-MH used $N_{i}=10$ inactive points and $10^{4}$
MCMC iterations; AS-PMMH used $N_{i}=10$ inactive points, 6 intermediate
distributions in the SMC used in the marginal likelihood estimates,
and 1666 MCMC iterations, with AS-PMMH-i using the corresponding configuration;
AS-MwG used $5\times10^{5}$ sweeps. The proposals we chosen in the
same way as in \ref{subsec:Performance-of-AS-MH}. Figure \ref{fig:Comparing-AS-MwG-with}
shows the error across 50 runs of each algorithm. We see that AS-MwG
offers superior performance to the other methods, fulfilling the promise
of using the AS to improve MCMC performance.

\subsubsection{Metropolis-within-particle Gibbs}

\begin{figure}

\subfloat[{Points from a long MCMC run for the posterior of the Gaussian mixture
model.\label{fig:Points-from-a}}]{\includegraphics[scale=0.4]{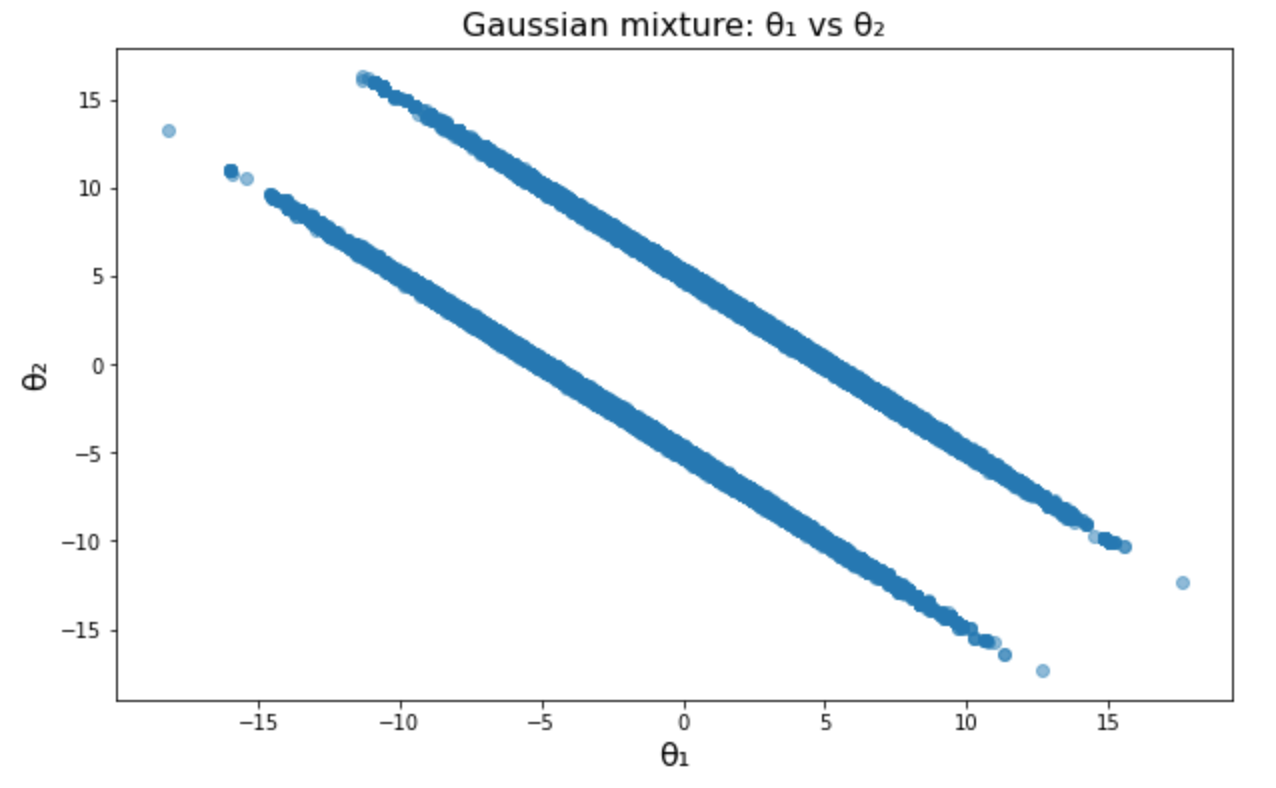}

}\subfloat[{The results of using MH on the Gaussian mixture model.\label{fig:The-results-of}}]{\includegraphics[scale=0.4]{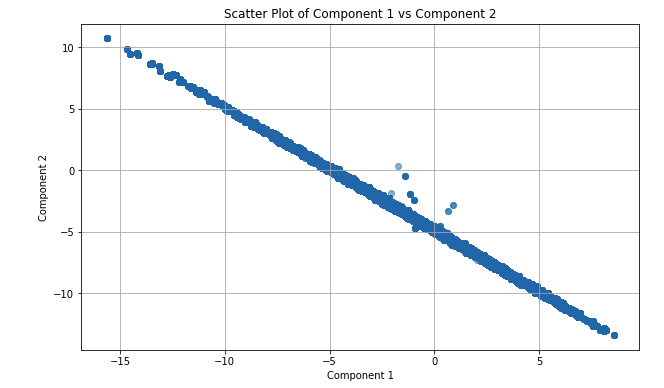}

}

\subfloat[{The results of using AS-MH on the Gaussian mixture model.\label{fig:The-results-of-1}}]{\includegraphics[scale=0.4]{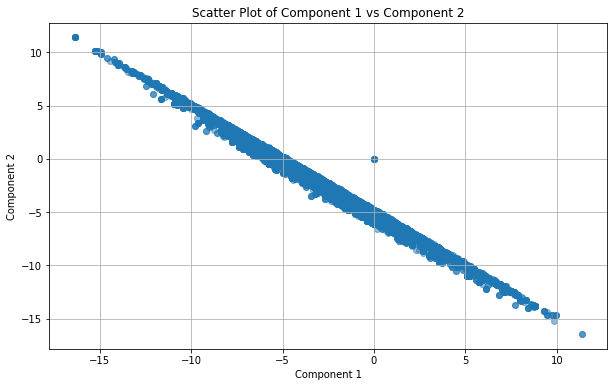}

}\subfloat[{The results of using AS-MwPG on the Gaussian mixture model.\label{fig:The-results-of-2}}]{\includegraphics[scale=0.4]{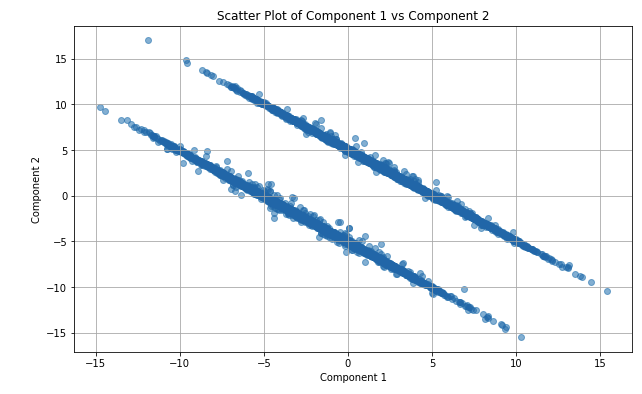}

}\caption{{Comparing MH, AS-MH and AS-MwPG on the mixture model.\label{fig:Comparing-MH,-AS-MH}}}

\end{figure}

To illustrate the use of AS-MwPG, we introduce a new example involving
a mixture model. We take the model for a single data point to be
\[
y\sim\frac{1}{2}\mathcal{N}\left(y\mid\theta_{1}+\theta_{2},1\right)+\frac{1}{2}\mathcal{N}\left(y\mid\theta_{3}+\theta_{4},1\right).
\]
We fit this model to 100 points drawn from $y\sim\frac{1}{2}\mathcal{N}\left(y\mid-5,1\right)+\frac{1}{2}\mathcal{N}\left(y\mid5,1\right)$,
using a multivariate Gaussian prior on $\theta$ with zero mean and
a diagonal covarisnce matrix with variance 25 on the diagonal. This
results in a posterior following the lines $\theta_{1}+\theta_{2}=\pm5$
and $\theta_{3}+\theta_{4}=\pm5$, as illustrated in figure \ref{fig:Points-from-a}.
For this example we find a two-dimensional active subspace. One of
the active and one of the inactive directions can be visualised in
figure \ref{fig:Points-from-a}, with the inactive subspace in the
direction parallel to the lines, and the active perpendicular to the
lines. We see that the challenge im this model is that there two modes
in the active direction, therefore we propose to use AS-MwPG so as
to use SMC for sampling the active variables.

We compare this method with MH and AS-MH, in all cases using a random-walk
Gaussian proposal on the active variables with (arbitrarily chosen)
variance $1/100$th of an estimated optimal covariance for the active
variables, in order to demonstrate that AS-MwPG can explore the two
modes without a carefully-chosen proposal. A total of $\sim10^{5}$
likelihood evaluations were used for all algorithms, using 1666 MCMC
iterations for AS-MH and AS-MwPG and 10 particles with 6 targets in
each SMC algorithm, compared with $10^{5}$ iterations of MH. The
results for each algorithm are shown in figures \ref{fig:Comparing-MH,-AS-MH}.
Only plots of $\theta_{1}$ and $\theta_{2}$ are shown: the results
for $\theta_{3}$ and $\theta_{4}$ are similar. We see that only
AS-MwPG is capable of exploring the two modes, illustrating a case
when it is helpful to use an SMC algorithm to simulate the active
variables.

\section{Discussion\label{sec:Conclusions}}

In this paper we have examined the performance of the AS-MH algorithm
of \citet{schuster_exact_2017}, which uses the idea of \citet{constantine_accelerating_2016}
in the context of an exact MCMC algorithm. To be computationally efficient,
these algorithms require a low variance IS estimate of the marginal
likelihood of active variables. However, we have seen that outside
of the ideal case where a linear reparameterisation results in variables
that are unaffected by the likelihood, these algorithms are unlikely
to be effective due to the IS estimator of the marginal likelihood
having a high variance. To address this issue, we suggest a PMMH variant
of AS-MH, using an SMC marginal likelihood estimator in place of the
IS estimator. However, we find this approach to be too computationally
expensive to be a useful approach. To develop effective MCMC algorithms
that make use of the active subspace, we instead propose Metropolis-within-Gibbs
algorithms, and show empirically the improved performance of these
approaches. We anticipate this approach being of use even outside
of the case of an ideal AS.

Despite the improved performance achieved by the new methods in this
paper, the effectiveness of AS-based approaches is still fundamentally
limited by the use of a global linear reparameterisation. To make
further progress in improving the efficiency of MCMC through the use
of reparameterisations based on identifying non-identifiability, we
recommend the use of non-linear reparametrisations, possibly via locally-linear
methods. Further, it might be worth investigating approaches that
avoid the dichotomous approach of categorising variables as either
active or inactive.

\backmatter

\bmhead{Acknowledgements}

Leonardo Ripoli's work was supported by EPSRC grant EP/L016613/1 (the Centre for Doctoral Training in the Mathematics of Planet Earth). Richard G. Everitt was supported by EPSRC grant EP/W006790/1 and NERC grant NE/T00973X/1. We thank Martyn Plummer for his comments on an early version of this work.

\begin{appendices}

\section{Active subspace Metropolis-Hastings\label{sec:Active-subspace-Metropolis-Hasti}}

To see how to use active subspaces within MCMC, we first write the
posterior on the new parameterisation.
\begin{eqnarray}
\pi_{a,i}\left(a,i\right) & \propto & p_{a,i}\left(a,i\right)l\left(B_{a}a+B_{i}i\right),\label{eq:as_posterior}
\end{eqnarray}
where $p_{a,i}\left(a,i\right)=p\left(B_{a}a+B_{i}i\right)$. Since
$B_{a}$ and $B_{i}$ are orthonormal there is no Jacobian associated
with this change in parameterisation. Throughout, as in \citet{constantine_accelerating_2016},
we assume that we have available the distributions $p_{a}$ and $p_{i}$
arising from the factorisation $p_{a,i}\left(a,i\right)=p_{a}\left(a\right)p_{i}\left(i\mid a\right)$:
specifically, that we can evaluate $p_{a}$ and $p_{i}$ pointwise,
and that we can simulate from $p_{i}\left(\cdot\mid a\right)$. \citet{constantine_accelerating_2016}
describes how to construct an approximate MCMC algorithm (in the style
of \citet{Alquier2016}) on the active variables through using a numerical
estimate of the (unnormalised) marginal distribution
\begin{eqnarray*}
\tilde{\pi}_{a}\left(a\right) & = & \int_{i}p\left(B_{a}a+B_{i}i\right)l\left(B_{a}a+B_{i}i\right)di\\
 & = & p_{a}\left(a\right)\int_{i}p_{i}\left(i\mid a\right)l\left(B_{a}a+B_{i}i\right)di
\end{eqnarray*}
at every iteration of the algorithm. The second term is the marginal
likelihood, as described in section \ref{subsec:A-pseudo-marginal-approach}.
It is this formulation on which we build in this paper; algorithm
\ref{alg:AS-MH} gives the active subspace MH (AS-MH) algorithm from
\citet{schuster_exact_2017}).

\begin{algorithm}
\caption{Active subspace Metropolis-Hastings}\label{alg:AS-MH}

Initialise $a^0$;\

\For {$n=1:N_i$}
{
	$i^{n,0} \sim q_{i} \left(\cdot \mid a^0 \right)$;\

	\[
	\tilde{w}^{n,0} = \frac{p_{i}\left(i^{n,0} \mid a^{0} \right)l\left( B_{a}a^0 +B_{i}i^{n,0}\right)}{q_{i}\left(i^{n,0}\mid a^0\right)};
	\]\
}

$u^{0} \sim\mathcal{M}\left( \left( w^{1,0}, ..., w^{N_i,0} \right) \right),$ where for $n=1:N_i$
\[
w^{n,0} = \frac{\tilde{w}^{n,0}}{\sum_{p=1}^{N_{i}} \tilde{w}^{p,0}};\
\]

Let $\bar{l}^0_{a}=\frac{1}{N_{i}}\sum_{n=1}^{N_{i}} \tilde{w}^{n,0}$;\

\For {$m=1:N_a$}
{
	$ a^{*m} \sim q_a\left(\cdot\mid a^{m-1} \right)$;\

	\For {$n=1:N_i$}
	{

		$i^{*n,m} \sim q_{i} \left(\cdot \mid a^{*m} \right);$\

		\[
		\tilde{w}^{*n,m} = \frac{p_{i}\left(i^{*n,m} \mid a^{*m} \right) l\left( B_{a} a^{*m} +B_{i} i^{*n,m}\right)}{q_{i}\left( i^{*n,m}\mid a^{*m} \right)};
		\]\
	}

	$u^{*m} \sim\mathcal{M}\left( \left( w^{*1,m}, ..., w^{*N_i,m} \right) \right),$ where for $n=1:N_i$
	\[
	w^{*n,m} = \frac{\tilde{w}^{*n,m}}{\sum_{p=1}^{N_{i}} \tilde{w}^{*p,m}};\
	\]

	Let $\bar{l}_{a}\left(a^{*m} \right)=\frac{1}{N_{i}}\sum_{n=1}^{N_{i}} \tilde{w}^{*n,m} $;\

    Set $\left(a^m, \left\{ i^{n,m}, w^{n,m} \right\}_{n=1}^{N_i}, u^m, \bar{l}^m_{a} \right) = \left(a^{*m},\left\{ i^{*n,m}, w^{*n,m} \right\}_{n=1}^{N_i}, u^{*m}, \bar{l}_{a}\left(a^{*m} \right) \right)$ with probability
	\[
	\alpha_a^m = 1\wedge\frac{p_{a}\left(a^{*m} \right) \bar{l}_{a}\left(a^{*m}\right)}{p_{a}\left(a^{m-1} \right) \bar{l}^{m-1}_{a}}\frac{q_a\left(a^{m-1} \mid a^{*m}\right)}{q_a\left(a^{*m}\mid a^{m-1} \right)};
	\]\
	
	Else let $\left(a^m, \left\{ i^{n,m}, w^{n,m} \right\}_{n=1}^{N_i}, u^m, \bar{l}^m_{a} \right) = \left(a^{m-1},\left\{ i^{n,m-1}, w^{n,m-1} \right\}_{n=1}^{N_i}, u^{m-1}, \bar{l}^{m-1}_{a}\right)$;\

}
\end{algorithm}

The output from algorithm \ref{alg:AS-MH} may be used to estimate
an integral
\[
\mathbb{E}_{\pi}\left[g\left(\theta\right)\right]=\int_{\theta}g\left(\theta\right)\pi\left(\theta\right)d\theta
\]
with respect to the posterior distribution $\pi$. We consider two
ways of composing estimators (following \citet{Andrieu2010c}).
\begin{enumerate}
\item \textbf{Using one $\theta_{i}$-point for each $\theta_{a}$-point:
\begin{equation}
\hat{\mathbb{E}}_{\pi}\left[g\left(\theta\right)\right]_{1}=\frac{1}{N_{a}}\sum_{m=0}^{N_{a}}g\left(B_{a}a^{m}+B_{i}i^{u^{m},m}\right)\label{eq:est1}
\end{equation}
}
\item \textbf{Using all of the accepted $\theta_{i}$-points for each $\theta_{a}$-point:}
\begin{equation}
\hat{\mathbb{E}}_{\pi}\left[g\left(\theta\right)\right]_{2}=\frac{1}{N_{a}}\sum_{m=0}^{N_{a}}\sum_{n=1}^{N_{i}}w^{n,m}g\left(B_{a}a^{m}+B_{i}i^{n,m}\right).\label{eq:est2}
\end{equation}
\end{enumerate}
\end{appendices}


\bibliography{smc_active_subspace}

\begin{thebibliography}{}
\providecommand{\doi}[1]{\url{https://doi.org/#1}}
\bibcommenthead

\bibitem[\protect\citeauthoryear{Alquier, Friel, Everitt, and Boland}{Alquier
  et~al.}{2016}]{Alquier2016}
Alquier, P., N.~Friel, R.G. Everitt, and A.~Boland. 2016.
\newblock Noisy {{Monte Carlo}}: {{Convergence}} of {{Markov}} chains with
  approximate transition kernels.
\newblock {\em Statistics and Computing\/}~{\em 26\/}(1): 29--47 .

\bibitem[\protect\citeauthoryear{Andrieu, Doucet, and Holenstein}{Andrieu
  et~al.}{2010}]{Andrieu2010c}
Andrieu, C., A.~Doucet, and R.~Holenstein. 2010.
\newblock Particle {{Markov}} chain {{Monte Carlo}} methods.
\newblock {\em Journal of the Royal Statistical Society: Series B\/}~{\em
  72\/}(3): 269--342 .

\bibitem[\protect\citeauthoryear{Andrieu and Roberts}{Andrieu and
  Roberts}{2009}]{Andrieu2009}
Andrieu, C. and G.O. Roberts. 2009.
\newblock The pseudo-marginal approach for efficient {{Monte Carlo}}
  computations.
\newblock {\em The Annals of Statistics\/}~{\em 37\/}(2): 697--725.
\newblock \doi{10.1214/07-AOS574} .

\bibitem[\protect\citeauthoryear{Beaumont}{Beaumont}{2003}]{Beaumont2003}
Beaumont, M.A. 2003.
\newblock Estimation of population growth or decline in genetically monitored
  populations.
\newblock {\em Genetics\/}~{\em 164\/}(3): 1139--1160 .

\bibitem[\protect\citeauthoryear{Beskos, Crisan, and Jasra}{Beskos
  et~al.}{2014}]{Beskos2014b}
Beskos, A., D.~Crisan, and A.~Jasra. 2014.
\newblock On the {{Stability}} of {{Sequential Monte Carlo Methods}} in {{High
  Dimensions}}.
\newblock {\em The Annals of Applied Probability\/}~{\em 24\/}(4): 1396--1445.
\newblock {\href{https://arxiv.org/abs/1103.3965v1}{{arxiv:1103.3965v1}}} .

\bibitem[\protect\citeauthoryear{Brouwer and Eisenberg}{Brouwer and
  Eisenberg}{2018}]{brouwer_underlying_2018}
Brouwer, A.F. and M.C. Eisenberg. 2018, February.
\newblock The underlying connections between identifiability, active subspaces,
  and parameter space dimension reduction.

\bibitem[\protect\citeauthoryear{Chopin}{Chopin}{2002}]{Chopin2002}
Chopin, N. 2002.
\newblock A sequential particle filter method for static models.
\newblock {\em Biometrika\/}~{\em 89\/}(3): 539--552.
\newblock \doi{10.1093/biomet/89.3.539} .

\bibitem[\protect\citeauthoryear{Constantine, Kent, and
  {Bui-Thanh}}{Constantine et~al.}{2016}]{constantine_accelerating_2016}
Constantine, P.G., C.~Kent, and T.~{Bui-Thanh}. 2016.
\newblock Accelerating {{Markov Chain Monte Carlo}} with {{Active Subspaces}}.
\newblock {\em SIAM Journal on Scientific Computing\/}~{\em 38\/}(5):
  A2779--A2805.
\newblock \doi{10.1137/15M1042127} .

\bibitem[\protect\citeauthoryear{Del~Moral, Doucet, and Jasra}{Del~Moral
  et~al.}{2006}]{DelMoral2006c}
Del~Moral, P., A.~Doucet, and A.~Jasra. 2006.
\newblock Sequential {{Monte Carlo}} samplers.
\newblock {\em Journal of the Royal Statistical Society: Series B\/}~{\em
  68\/}(3): 411--436 .

\bibitem[\protect\citeauthoryear{Douc, Capp{\'e}, and Moulines}{Douc
  et~al.}{2005}]{douc_comparison_2005}
Douc, R., O.~Capp{\'e}, and E.~Moulines 2005.
\newblock Comparison of resampling schemes for particle filtering.
\newblock In {\em {{ISPA}} 2005. {{Proceedings}} of the 4th {{International
  Symposium}} on {{Image}} and {{Signal Processing}} and {{Analysis}}, 2005.},
  {Zagreb, Croatia}, pp.\  64--69. {IEEE}.

\bibitem[\protect\citeauthoryear{Drton and Plummer}{Drton and
  Plummer}{2017}]{drton_bayesian_2017}
Drton, M. and M.~Plummer. 2017, March.
\newblock A {{Bayesian Information Criterion}} for {{Singular Models}}.
\newblock {\em Journal of the Royal Statistical Society Series B: Statistical
  Methodology\/}~{\em 79\/}(2): 323--380.
\newblock \doi{10.1111/rssb.12187} .

\bibitem[\protect\citeauthoryear{Everitt}{Everitt}{2024}]{everitt2024ensemble}
Everitt, R.G. 2024.
\newblock Ensemble kalman inversion approximate bayesian computation.
\newblock {\em arXiv preprint arXiv:2407.18721\/} .

\bibitem[\protect\citeauthoryear{Gill, Koskela, Didelot, and Everitt}{Gill
  et~al.}{2023}]{gill2023bayesian}
Gill, A., J.~Koskela, X.~Didelot, and R.G. Everitt. 2023.
\newblock Bayesian {{Inference}} of {{Reproduction Number}} from {{Epidemic}}
  and {{Genetic}} data using {{Particle MCMC}}.
\newblock {\em arXiv preprint arXiv:2311.09838\/} .

\bibitem[\protect\citeauthoryear{Kong, Liu, and Wong}{Kong
  et~al.}{1994}]{Kong1994}
Kong, A., J.S. Liu, and W.H. Wong. 1994.
\newblock Sequential {{Imputations}} and {{Bayesian Missing Data Problems}}.
\newblock {\em Journal of the American Statistical Association\/}~{\em
  89\/}(425): 278--288.
\newblock \doi{10.2307/2291224} .

\bibitem[\protect\citeauthoryear{Pitt, Silva, Giordani, and Kohn}{Pitt
  et~al.}{2012}]{Pitt2012}
Pitt, M.K., S.~Silva, P.~Giordani, and R.~Kohn. 2012.
\newblock On some properties of {{Markov}} chain {{Monte Carlo}} simulation
  methods based on the particle filter.
\newblock {\em Journal of Econometrics\/}~{\em 171\/}(2): 134--151.
\newblock \doi{10.1016/j.jeconom.2012.06.004} .

\bibitem[\protect\citeauthoryear{Schuster, Constantine, and Sullivan}{Schuster
  et~al.}{2017}]{schuster_exact_2017}
Schuster, I., P.G. Constantine, and T.J. Sullivan. 2017.
\newblock Exact active subspace {{Metropolis-Hastings}}, with applications to
  the {{Lorenz-96}} system.

\bibitem[\protect\citeauthoryear{Sherlock, Thiery, Roberts, and
  Rosenthal}{Sherlock et~al.}{2015}]{Sherlock2015}
Sherlock, C., A.H. Thiery, G.O. Roberts, and J.S. Rosenthal. 2015, February.
\newblock On the efficiency of pseudo-marginal random walk {{Metropolis}}
  algorithms.
\newblock {\em The Annals of Statistics\/}~{\em 43\/}(1): 238--275.
\newblock \doi{10.1214/14-AOS1278}.
\newblock {\href{https://arxiv.org/abs/1309.7209v3}{{arxiv:1309.7209v3}}} .

\end{thebibliography}

\end{document}